\documentclass[twocolumn,twocolappendix,]{emulateapj}
\usepackage{apjfonts}

\newcommand{\Msun}{$M_{\sun}$}

\usepackage{color}
\definecolor{lightred}     {rgb}{1.00, 0.35, 0.05}
\definecolor{darkred}     {rgb}{0.70, 0.05, 0.05}
\definecolor{lightblue}    {rgb}{0.05, 0.35, 1.00}
\definecolor{darkblue}    {rgb}{0.05, 0.05, 0.70}

\usepackage{amsmath}

\usepackage{natbib}
\usepackage{graphicx}

\def\s4g{{S$^4$G}}
\def\Msun{$M_\odot$}
\def\ser{S{\'e}rsic}
\def\mus{{{$\mu m$ }}}
\def\micron{{{$\mu m $}}}
\def\mum{{{$\mu m$}}}
\shorttitle{Mass profile and global shape of bars in {\s4g}}
\shortauthors{Kim et al.}

\begin{document}
\title{The Mass Profile and Shape of Bars in the {\em Spitzer}  Survey of Stellar Structure in Galaxies (S$^{4}$G): Search for an Age Indicator for Bars}

\author{
Taehyun Kim\altaffilmark{1,2,3},
Kartik Sheth\altaffilmark{2},
Dimitri A. Gadotti\altaffilmark{3},
Myung Gyoon Lee\altaffilmark{1},
Dennis Zaritsky\altaffilmark{4},
Bruce G. Elmegreen\altaffilmark{5},\\
E. Athanassoula\altaffilmark{6},
Albert Bosma\altaffilmark{6},
Benne Holwerda\altaffilmark{7},
Luis C. Ho\altaffilmark{8,9},
S\'ebastien Comer\'on\altaffilmark{10,11},
Johan H. Knapen\altaffilmark{12,13},
Joannah L. Hinz\altaffilmark{14},
Juan-Carlos Mu\~noz-Mateos\altaffilmark{2,3},
Santiago Erroz-Ferrer\altaffilmark{12,13},
Ronald J. Buta\altaffilmark{15},
Minjin Kim\altaffilmark{16},\\
Eija Laurikainen\altaffilmark{10,11},
Heikki Salo\altaffilmark{10},
Barry F. Madore\altaffilmark{17},
Jarkko Laine\altaffilmark{10},
Kar\'in Men\'endez-Delmestre\altaffilmark{18},\\
Michael W. Regan\altaffilmark{19},
Bonita de Swardt\altaffilmark{20},
Armando Gil de Paz\altaffilmark{21},
Mark Seibert\altaffilmark{17},
Trisha Mizusawa\altaffilmark{2,22}
}

\altaffiltext{}{\it Affiliations can be found before the references.}

\begin{abstract}
We have measured the radial light profiles and global shapes of bars using two-dimensional 3.6 {\micron} image decompositions for 144 face-on barred galaxies from the {\em Spitzer}  Survey of Stellar Structure in Galaxies (\s4g). 
The bar surface brightness profile is correlated with the stellar mass and bulge-to-total (B/T) ratio of their host galaxies.
Bars in massive and bulge-dominated galaxies (B/T$>$0.2) show a flat profile, while bars in less massive, disk-dominated galaxies (B/T$\sim$0) show an exponential, disk-like profile with a wider spread in the radial profile than in the bulge-dominated galaxies.    
The global two-dimensional shapes of bars, however, are rectangular/boxy, independent of the bulge or disk properties.  
We speculate that because bars are formed out of disk, bars initially have an exponential (disk-like) profile which evolves over time, trapping more stars into the boxy bar orbits. This leads bars to become stronger and have flatter profiles. 
The narrow spread of bar radial profiles in more massive disks suggests that these bars formed earlier (z$>$1), while the disk-like profiles and a larger spread in the radial profile in less massive systems imply a later and more gradual evolution, consistent with the cosmological evolution of bars inferred from observational studies.
Therefore, we expect that the flatness of the bar profile can be used as a dynamical age indicator of the bar to measure the time elapsed since the bar formation.
We argue that cosmic gas accretion is required to explain our results on bar profile and the presence of gas within the bar region.
\end{abstract}

\keywords{galaxies: evolution -- galaxies: formation -- galaxies: spiral -- galaxies: structure}

\section{Introduction}
The presence and properties of galactic structures such as bars, bulges, rings, and spiral arms are not pre-determined at the time of the galaxy formation.  Both fast and slow  (``secular'') processes can create galactic structures and change their properties by a rearrangement of the mass, angular momentum, and energy with time (\citealt{athanassoula_13_book, sellwood_13_rev}, also see reviews in \citealt{falcon_barroso_13_book}). 
As the merger rate in the Universe declines, the evolution of galaxies at their late stages has increasingly been governed by secular evolution (\citealt{kormendy_04}), stimulated by non-axisymmetric structures such as bars, ovals, spiral structures, triaxial dark matter halos. Amongst these, bars are one of the most important drivers of internal secular evolution in their host galaxies.

The non-axisymmetric potential of a bar induces large scale streaming motions in stars and gas into the central part of the galaxy (e.g., \citealt{athanassoula_92a, athanassoula_92b, sellwood_93}). 
Being dissipative, the gas loses angular momentum and energy and flows inwards towards the galactic center (\citealt{knapen_95a, regan_99, sheth_00, sheth_02}), accumulating in the central $\sim$kpc of galaxies (e.g., \citealt{sakamoto_99, sheth_05}).
The accumulation of gas in the central parts leads to high levels of circumnuclear star formation activity (\citealt{sersic_65, hawarden_86, devereux_87, martin_95, ho_97, sheth_00, sheth_05, gadotti_01, ellison_11, coelho_11, wang_12}); this circumnuclear star formation is often in the shape of nuclear rings (\citealt{knapen_02, comeron_10, kim_w_12c, seo_13}) and nuclear star-clusters (\citealt{boker_02, boker_04, boker_11}). Such star formation activities may help to  create disky pseudo bulges (\citealt{kormendy_04, sheth_05, athanassoula_05a, debattista_06}).  
Bars are the primary mechanism for transporting gas from a few kpc scale to the central kpc. 
However, there have been mixed answers to the question whether the presence of a bar and AGN activity are connected. Some studies find weak statistical links among AGN activity and the presence of bars (e.g., \citealt{arsenault_89, knapen_00, laine_02, laurikainen_04_bar, coelho_11}), whereas others find little or no link (e.g., \citealt{moles_95, mcleod_95, mulchaey_97, ho_97, hunt_99, martini_03b, lee_12, cisternas_13}).
While bar torques drive gas inside the bar co-rotation inwards, they push the gas between the co-rotation and outer Lindblad resonance (OLR) outwards (\citealt{combes_08_gas, kubryk_13}).   

Bars have been reported to change the chemical abundance gradient in the disk, presumably due to large scale streaming motions induced by the bar (e.g., \citealt{zaritsky_94, martin_94, dutil_99}, but also see \citealt{perez_11, sanchez_12, sanchez_14}). 
Bars may change the chemical abundance gradient inside the co-rotation radius but they seem to have only a small impact outside the bar itself.  Bars may redistribute stars in the galaxy disk leading to disk breaks in the disk profile (e.g. \citealt{debattista_06, radburn_smith_12, munoz_mateos_13, kim_14a}, hereafter Paper I).  
Bars drive the formation of inner rings and outer rings (\citealt{buta_96, buta_03, romero_gomez_06, athanassoula_09a, athanassoula_09b}), and possibly  spiral density waves (\citealt{buta_09, salo_10}).  Simulations show that bars may evolve over time by transferring angular momentum from the baryons to the outer disk and/or halo (\citealt{sellwood_80, debattista_98, athanassoula_02b, athanassoula_03, valenzuela_03, martinez_valpuesta_06, begelman_09, saha_12}). As bars lose angular momentum, their corotation radius moves outwards, and they become longer and thinner (\citealt{athanassoula_03, athanassoula_13_book}).  
Bars thus play an important role in secular evolution of galaxies by redistributing the energy, angular momentum and mass across the disk.   

Bar properties change along the Hubble sequence. Early Hubble type disks (earlier than SBb) have longer bars, both in an absolute sense and relative to their disks, compared to late Hubble type disk galaxies 
(later than SBb, \citealt{elmegreen_85,martin_95, laurikainen_02_2mass, laurikainen_04_bar, erwin_05_bar, laurikainen_07, menendez_delmestre_07, marinova_07}).
Bars in early type spirals tend to show uniform intensity along the major axis of the bar, i.e., a flat radial light profile  compared to the inter-bar region, whereas bars in late type spirals tend to have exponential radial profiles (\citealt{elmegreen_85, baumgart_86, elmegreen_96a, regan_97}).  Flat bars are associated with two-arm grand design spirals. However, exponential bars have multiple spirals or flocculent spirals (\citealt{elmegreen_85}), and these spirals are often not physically connected directly to the bar structure. 
As bars evolve, stellar orbits of bars also evolve (\citealt{athanassoula_13_book}). Thus they may be different between early and late type disk galaxies. Such orbits define the outermost two-dimensional (2D) shape of bars. Therefore if we investigate both bar profile and shape over Hubble type, and as a function of structural properties of galaxies, we should be able to better understand how bars evolve.

Previous studies have analyzed at most a few dozen bars with relatively simple (one-dimensional) analysis of their light profile. Although some galaxies have been analyzed using 2D decompositions including bar components (e.g., \citealt{laurikainen_04_osubg, laurikainen_11, gadotti_09, vika_12}), properties of bars have not yet been studied in detail.
With the {\em Spitzer}  Survey of Stellar Structure in Galaxies (S4G, Sheth et al. 2010), we now have the opportunity to measure the bar light profile, the bar shape, and bulge and disk properties using the survey of 3.6 {\micron} images, which are less affected by dust, in a sizable sample of galaxies shedding light on the evolution of bars and disks. The large, uniform and homogenous 3.6 {\mus} data give us a virtually dust-free view of stellar structures which is important because dust can affect the measurement of galactic structures (e.g., \citealt{holwerda_05_iv, gadotti_10, fathi_10a, kelvin_12, Pastrav_13}).   

As bars and bulges are embedded in disks, structural properties of bars are best studied through 2D decompositions (e.g., \citealt{dejong_96II, laurikainen_05, laurikainen_07, laurikainen_10, gadotti_08, gadotti_09_SDSS, durbala_08}, Paper I).  
A shortcoming for most of the previous studies has been the use of a fixed profile for the bar and of a single exponential for the disk. 
Although several studies have tried to fit bars with {\ser} or Ferrers function (e.g., \citealt{laurikainen_05, laurikainen_10,weinzirl_09, vika_12, lansbury_14}), light profile of bars have not been examined.
Moreover, a majority of disk galaxies have a disk break (\citealt{pohlen_02, erwin_05_anti, pohlen_06, erwin_08, gutierrez_11, maltby_12a, comeron_12, martin_navarro_12, munoz_mateos_13, laine_14}), therefore it is critical to fit both the inner and outer disks (Paper I).  The disk break affects the measurement of the  structural properties of  the bar, bulge and the disk. For example in down-bending (Type II) disks, we find that B/T and bar-to-total (Bar/T) can change up to 10\% and 25\%, respectively (Paper I).   The disk scale length and central surface brightness of the disk also change once the disk break is properly modeled.  However, none of the previous studies consider disk breaks in measuring structural properties of galaxies.
In this work we allow the bar profile to vary and fit the disk break in 144 galaxies from \s4g. The radial profile and global shape of bars are analyzed with respect to the bulge and disk properties with an aim of understanding the evolution of disks.

The paper is organized as follows. In \S 2 we give a brief overview of our sample selection and describe our 2D image decomposition procedure. We explore the radial surface brightness profile of the bar in \S 3. Global shapes of bars are examined in \S 4.  
We discuss our results in \S 5, and summarize our results and conclude in \S 6.

\section{Data and Data Analysis}
We refer the reader to Paper I for details on the sample selection and the 2D decomposition methodology, and briefly summarize the data set and analysis here.  

\subsection{Data}
We use data from the {\em Spitzer} Survey of Stellar Structure in Galaxies ($S^4G$, \citealt{sheth_10}), a survey of 2,352 nearby galaxies using the Infrared Array Camera (IRAC, \citealt{fazio_04}) at 3.6 and 4.5 \mum. 
Mid-infrared (MIR) data is a good tracer of the stellar mass distribution in galaxies, because it is less hampered by dust with only a small local contamination from AGB stars or hot dust around red super giants (\citealt{meidt_12a, meidt_12b}). Thus in this study, we made use of 3.6 {\mus} images that form an ideal data set for exploring properties of stellar bars.


Our sample of 144 barred galaxies are all the barred galaxies from the data that had been processed by the \s4g pipelines (Pipelines 1, 2, and 3; \citealt{sheth_10}) at the beginning of this study in November 2011. Properties of galaxies are presented in the Paper I. The bar classification was first done visually by K. Sheth, T. Kim, and B. de Swardt, and then later the presence of a bar was confirmed by comparing with the MIR classification (Buta et al. 2010, R. Buta et al. 2014, submitted to ApJS). According to this MIR classification, there are $\sim$ 800 stongly barred (SB) galaxies, and $\sim$ 370 weakly barred (SAB) galaxies in the full sample of the \s4g.

\subsection{Data Analysis}
\label{sec:data}
We performed 2D decompositions on 3.6 \mus images from \s4g using the BUlge Disk Decomposition Analysis code ({\sc BUDDA v2.2}, \citealt{gadotti_08}, \citealt{desouza_04}) and fit each galaxy with a disk, bar, bulge, and if present, a central point source.  
As noted earlier, the majority of disks have a change of slope in their radial light distribution with either down-bending (Type II) or up-bending (Type III) profiles (\citealt{pohlen_06, hunter_06, erwin_08, gutierrez_11, maltby_12b, munoz_mateos_13, laine_14}). Due to the disk breaks, disks have inner and outer disk scale lengths which differ in the median by 40\% (Paper I) and lead to differences of $\sim$10\% in B/T, and $\sim$25\% in Bar/T in the decompositions.   

We follow the procedures detailed in the Paper I, but summarize the fitting procedues of bars here.
The bar surface brightness profile is also modeled with the {\ser} profile (\citealt{sersic_63}).
\begin{equation}
\mu_{\rm bar}(r)= \begin{cases}
\mu_{e,{\rm bar}}+c_{n,{\rm bar}}\left[\left(\frac{r}{r_{e,{\rm bar}}}\right)^{1/n_{{\rm bar}}}-1\right], & \mbox  {if } r\leq r_{\rm bar}
\\
0, &   \mbox {if } r > r_{\rm bar}
\end{cases}
\end{equation}
\noindent where $c_{n,{\rm bar}}=2.5(0.868n_{{\rm bar}}-0.142)$.
$r_{{\rm bar}}$ is the radius of the bar along the major axis.
Beyond this radius the light profile of the bar is truncated to zero level in the model images.

The global shape of each component can be modeled with generalized ellipses (\citealt{athanassoula_90}),
\begin{equation}
\left( \left| x \right| \over a \right)^c + \left( \left| y \right| \over b \right)^c  = 1, 
\end{equation} \label{eq_bar_shape}
where $x$ and $y$ are position of points, $a$ and $b$ are semi major and semi minor axis, respectively, and $c$ is the shape parameter which describes the shape of the ellipse.
If $c=2$, then the shape of the component is a perfect ellipse. If $c<2$, then the shape of the component is disky while if the $c>2$, the shape of the component is boxy. In this study, we only characterize the shape of a bar component, and we fix the shape parameter ($c=2$) for disk and bulge.

In case a nuclear point source is present (23/144, 16\% of the sample), we model it with a PSF profile with its approriate FWHM, while we fit only for the peak intensity. 
Possible candidates of nuclear sources are non stellar emission from active galactic nuclei, nuclear star clusters or unresolved small bulges. 
The lowest B/T that we obtain is 0.004. This is the lower limit that we can identify a bulge visually from our analysis.
In our sample, there are a number of bulgeless galaxies. Before we run {\sc budda}, we visually inspect the images and radial light profiles of the galaxies and if a galaxy does not have a bulge, then we only fit these galaxies with a disk and a bar, thus such galaxies have B/T$=$0.
In this study, by ``bulgeless'' galaxy, we refer to the galaxies without a classical nor a disky pseudo bulge. But still bulgeless galaxies can have a boxy/peanut feature, which is sometimes called boxy-peanut bulge (For details, readers are refered to \citealt{athanassoula_05a, athanassoula_06}).
\begin{figure}
\begin{center}
\includegraphics[width=8cm]{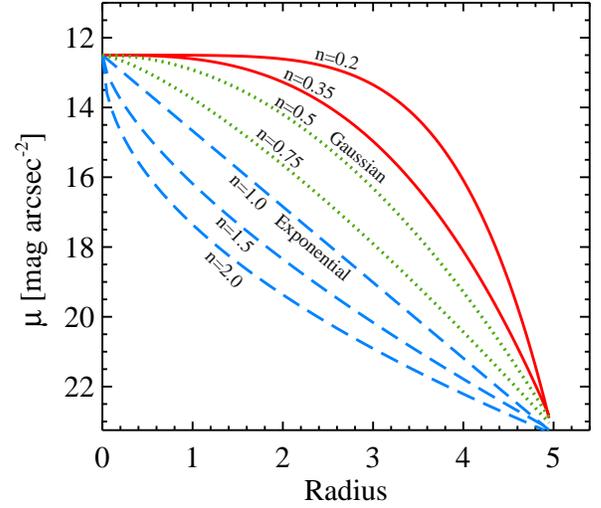}
\caption{Schematic plot of {\ser} profiles which describe various radial surface brightness profiles of bars. Radial profiles of n$=$0.2 to 2 are presented. Gaussian (n$=$0.5) and Exponential (n$=$1.0) profiles are also shown. Flat profiles are in red, intermediate profiles are in green, and exponential steep profiles are in blue. Radius is in an arbitrary unit.}
\label{fig:n_bar}
\end{center}
\end{figure}

\begin{figure}
\begin{center}
\includegraphics[width=8cm]{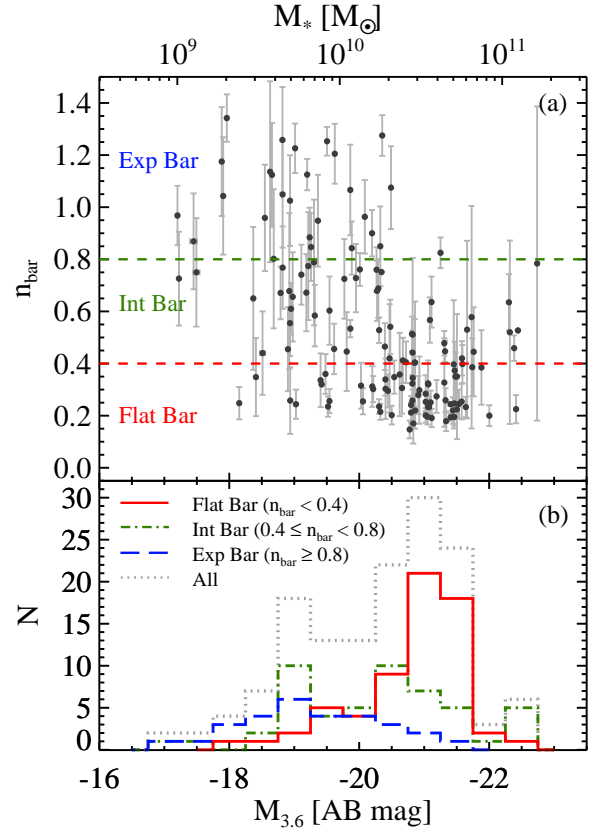}
\caption{(a): Bar {\ser} indices as a function of absolute magnitude of galaxies. Stellar masses, converted from absolute magnitude of galaxies, are shown on top. Error bars are statistical 1$\sigma$ uncertainty errors for {\ser} indices.
(b): Distributions of absolute magnitude of galaxies having flat bars ($n_{\rm bar} <$ 0.4) in red solid line, intermediate bars (0.4 $\le n_{\rm bar} <$ 0.8) in green dot-dahsed line, exponential bars ($n_{\rm bar} \ge$0.8) in blue dashed line.}
\label{fig:n_bar_abmag}
\end{center}
\end{figure}

\section{Surface Brightness Profile of Bars}
\label{sub_bar_prof}
\subsection{Bar Profiles Fitted with {\ser} Function}\label{subsec:bar_profile}

We use {\ser} indices to measure the steepness of the light profile of a bar. Surface brightness profiles with different {\ser} indices from n$=$0.2 to 2.0 are shown in Fig~\ref{fig:n_bar}.  
In Fig~\ref{fig:n_bar_abmag}(a), we plot bar {\ser} indices ($n_{\rm {bar}}$) as a function of the absolute magnitude of the galaxy. The 3.6 {\micron} magnitudes are converted to a stellar mass  following the method outlined in Appendix A of \citet{munoz_mateos_13}, which is based on the mass-to-light ratio measurement from \cite{eskew_12}. 
Fig~\ref{fig:n_bar_abmag}(b) shows the distributions of galaxies that have a flat bar, an intermediate bar, and an exponential bar.
We find that massive galaxies predominantly have flat bars ($n_{\rm {bar}}<0.4$), while less massive galaxies primarily have exponential ``disk-like'' bars ($n_{\rm {bar}} \ge 0.8$), although some low mass galaxies show flat bars.
The transition from having predominantly flat bars to more exponential bars occurs around $M_{3.6} \sim -20$ AB mag ($\sim M_{\ast}$/{\Msun}$\sim 10^{10.2}$).  

\begin{figure*}
\begin{center}
\includegraphics[width=\textwidth]{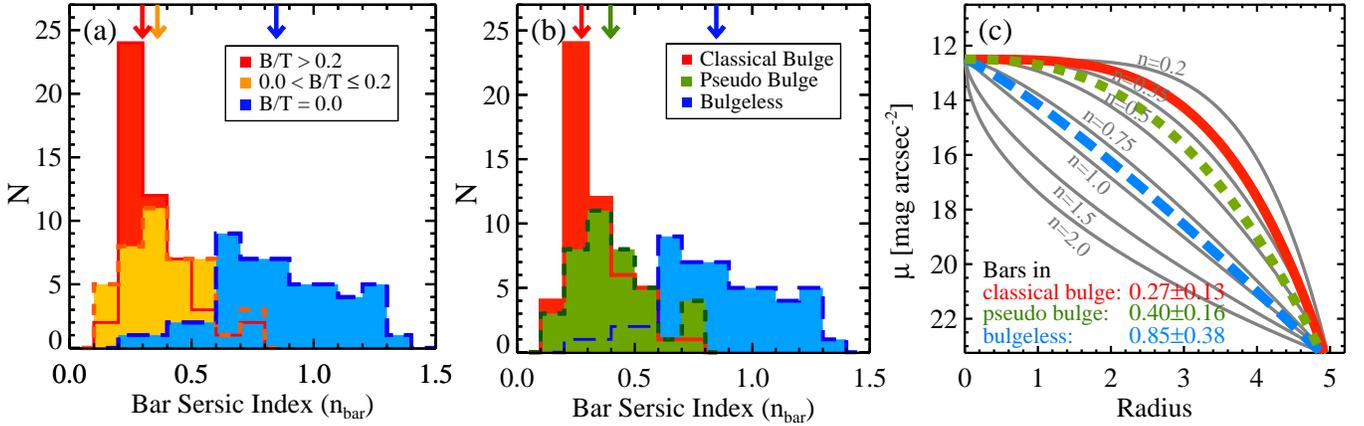}
\caption{Distribution of bar {\ser} indices. (a) The sample is divided into three groups, galaxies with B/T$>$0.2 , 0$<$B/T$\leq$0.2, and B/T$=$0. Downward arrows indicate median bar {\ser} indices for each group. (b) The sample is split into galaxies with a classical bulge, disky pseudo bulge, and bulgeless galaxies. 
Galaxies which have a bulge component have smaller bar {\ser} indices than bulgeless galaxies do which implies that galaxies which have a bulge have flatter bars than bulgeless galaxies do. (c) Median bar profiles of classical bulge galaxies (red solid line), disky pseudo bulge galaxies (green dashed line), and bulgeless galaxies (blue short-dashed line) are plotted. Median {\ser} indices are also shown on the bottom.}
\label{fig:n_bar_hist}
\end{center}
\end{figure*}

Next we plot the distribution of $n_{\rm {bar}}$ in Fig~\ref{fig:n_bar_hist}(a) and (b) dividing the galaxies now based on the bulge-dominance: in red we show the bulge-dominated galaxies with a B/T\footnote{B/T is from our own 2D decompositions using 3.6 {\micron} images.} $>$ 0.2, intermediate cases are shown in orange where 0.0 $<$ B/T $\leq$ 0.2, and disk-dominated systems are shown in blue with B/T $=$ 0.
The arrows at the top of the panel indicate the median $n_{\rm {bar}}$ for each group.
The main point to note is that the distributions of $n_{\rm {bar}}$ of the three groups are different.
Bulge-dominated galaxies have a smaller $n_{\rm {bar}}$ with a median bar {\ser} index, $<n_{\rm {bar}}>$ $\sim $ 0.30, whereas disk-dominated systems span a wide range of bar {\ser} index from 0.25 to 1.4, with a median $<n_{\rm {bar}}> \sim$ 0.85. The majority of exponential bars are in bulgeless galaxies, and all galaxies with $n_{\rm {bar}}$ > 0.7 are bulgeless galaxies. Thus bar profiles can be better separated by bulge dominance and bulge types than by galaxy mass.

Next we investigate the bar {\ser} index versus the bulge {\ser} index. We divide the bulge light profiles into three groups:   
no bulge ($n_{\rm bulge} =$ 0), disky pseudo bulge\footnote{In this paper, by ``disky pseudo bulge'', we specifically refer to disk-like or disky bulge, and do not include boxy peanut features that are thick part of the bar. See \citet{athanassoula_05a, athanassoula_06} for details.} (0 < $n_{\rm bulge} \leq$ 2.0), and a classical bulge ($n_{\rm bulge}$ > 2.0) following the separation of \citet{fisher_08}. 

The bar {\ser} index distribution for these groups is shown in Fig~\ref{fig:n_bar_hist}(b).  We find that classical bulge galaxies have a smaller $n_{\rm bar}$ and show a flatter bar profile compared to bulgeless galaxies. In Fig~\ref{fig:n_bar_hist}($c$) we plot the different bar profiles but now we overplot the median bar profiles for galaxies with a classical bulge with the red solid line, galaxies with a disky pseudo bulge with the green dashed line, and bulgeless galaxies with the blue short-dashed line.  There does seem to be a progression of $n_{\rm bar}$ but the samples are still not large enough to get a statistically significant differences in the distribution of the bar profiles between the classical and disky pseudo bulge groups. Nevertheless the basic conclusion from these figures is that the bulge-dominated galaxies and especially those with a classical bulge have the flattest bars, and this is consistent with the early findings of \citet{elmegreen_85}.

\begin{figure}
\begin{center}
\includegraphics[width=8cm]{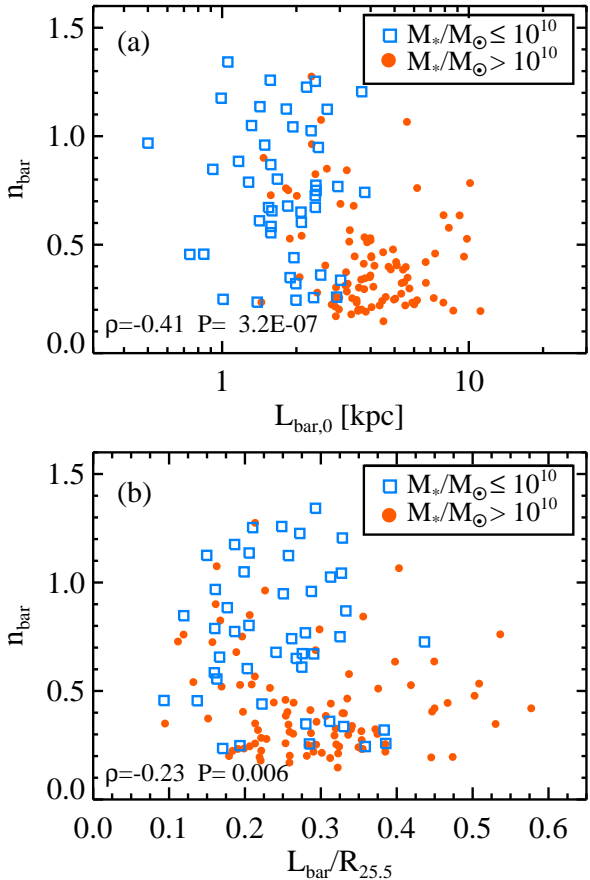}
\caption{(a): Bar {\ser} indices ($n_{\rm bar}$) with bar lengths ($L_{\rm bar}$) and (b): ($n_{\rm bar}$) with normalized bar lengths ($L_{\rm bar}$/$R_{\rm 25.5}$). Massive galaxies are plotted in circles while less massive galaxies are in squares. Spearman's rank correlation coefficient ($\rho$) and significance are presented on the lower left corner of each panel.}
\label{fig:n_bar_l_bar}
\end{center}
\end{figure}

We check how $n_{\rm bar}$ vary with bar length ($L_{\rm bar}$) and normalized bar length ($L_{\rm bar}$/ $R_{\rm 25.5}$\footnote{
Radius at $\mu_{3.6 \mu m} = $25.5 AB mag arcsec$^{-2}$ from the \s4g Pipeline 3.}) in Fig~\ref{fig:n_bar_l_bar}. 
We find that longer bars tend to show flatter profiles (Fig~\ref{fig:n_bar_l_bar}a). This is in line with the previous studies (e.g., \citealt{elmegreen_85, baumgart_86}), considering that early type galaxies have longer bars (e.g., \citealt{erwin_05_bar, laurikainen_07,menendez_delmestre_07}). 
However, when lengths of bars are normalized to $R_{\rm 25.5}$, the correlation is not clear (Fig~\ref{fig:n_bar_l_bar}b).

For highly inclined galaxies, projection effects could be such that bulge light is mixed with the bar, complicating the decomposition. Nevertheless, this effect is unlikely to influence much in our sample, because we select our samples to have mild inclination ($i<60$). The mean bulge effective radius to bar radius is 7.5 for galaxies with a bulge. Thus bars span large enough area compared to bulges, enabling us to well estimate bar {\ser} indices.
\subsection{Fitting Bar Model Images with the Modified Ferrers Profile}
\begin{figure*}
\begin{center}
\includegraphics[width=\textwidth]{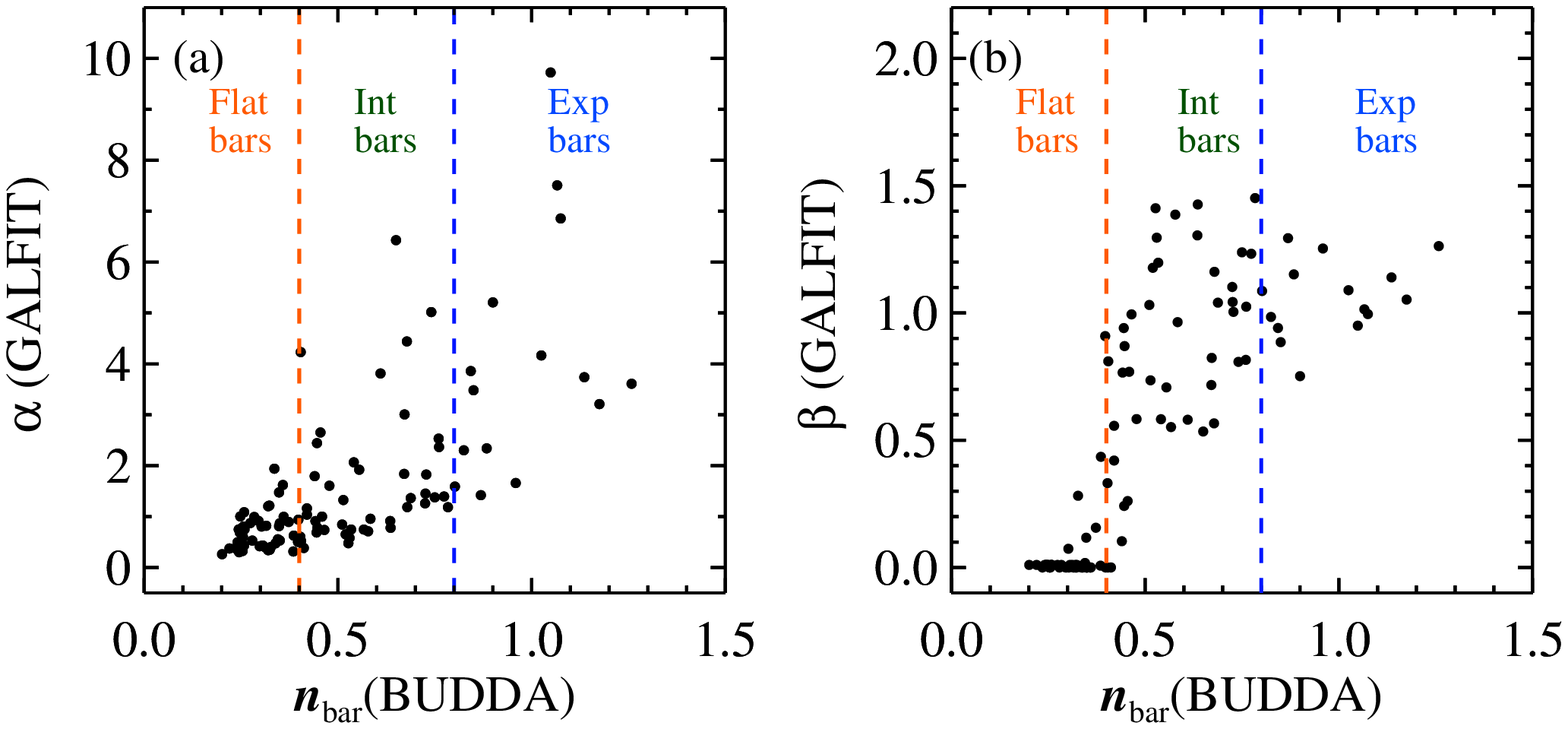}
\caption{Comparisons of bar radial profile parameters. Bar {\ser} index ($n_{\rm bar}$) from the {\ser} profile fit,  and $\alpha$ and $\beta$ from the modified Ferrers function (Eq.~\ref{eq:ferrer}).}
\label{fig:bar_ser_ferrer}
\end{center}
\end{figure*}

We model the bar profile with the {\ser} profile that is provided by {\sc budda} to fit bars.
However, bars also have been modeled with the Ferrers function (e.g., \citealt{laurikainen_07, laurikainen_10, peng_10}). 
The Ferrers function has the following functional form:
\begin{equation}
\mu_{\rm bar}(r)=\mu_0 (1 - (r/r_{\rm out})^{2})^{\rm m_{\rm bar}+0.5}, \\
\label{eq:ferrer_original}
\end{equation}
where $\rm m_{\rm bar}$ is a parameter that defines the shape of bar profiles. The function is only defined out to $r_{\rm out}$, which is the bar radius. Beyond $r_{\rm out}$, $\rm \mu_{\rm bar}$ is set to 0. 
\cite{kim_w_12c} compared their hydrodynamical simulations of galaxies that have a bulge with the observational study of \citet{comeron_10}{\footnote{90\% of their sample galaxies have Hubble T$\leq$4 (SBbc).}}, and show that observed bars are best represented by $m_{\rm bar} \leq$ 0.5. 

It would be instructive to compare the two functions ({\ser} and Ferrers profile) for the fits.
However, it is not straightforward to convert {\ser} indices to the $m_{\rm bar}$. Therefore to compare $n_{\rm bar}$ and $m_{\rm bar}$, we ran GALFIT (version 3.0.5 Peng et al. 2010, 2002) on the bar model images that were obtained with  {\sc budda} in order to estimate parameters of the Ferrers profile.
{\sc GALFIT} presents the modified Ferrers profile that has the following functional form:
\begin{equation}
\mu_{\rm bar}(r)=\mu_0 (1 - (r/r_{\rm out})^{2-\beta})^{\alpha}, \\
\label{eq:ferrer}
\end{equation}
where $\mu_0$ is the central surface brightness, $\alpha$ describes how sharply the bar profile drops near $r_{\rm out}$, and $\beta$ describes the inner central slope of the profile. Because of its ability to describe a flat core and sharp truncation, the modified Ferrers profile is often used to model bars or lenses (\citealt{peng_10}). We refer the readers to Fig 4 of \citet{peng_10} for details about the profiles.

Parameters of the Ferrers function and modified Ferrers function are different in a way that $\alpha =  m_{\rm bar} - 0.5$, and $\beta$ is fixed to 0 in the Ferrers function. But as we will see in Fig~\ref{fig:bar_ser_ferrer}, many of our bar models, especially those that have no bulge, do not have $\beta=$0. 

We compare $\alpha$ and $n_{\rm bar}$, and $\beta$ and $n_{\rm bar}$ in Fig~\ref{fig:bar_ser_ferrer}. For 35 of the 144 galaxies GALFIT did not converge to a meaningful solution and for those we were not able to obtain $\alpha$ and $\beta$ values.
As we expected, $n_{\rm bar}$ and $\beta$ are strongly related, and $n_{\rm bar}$ and $\alpha$ also show a correlation. Flat bars ($n_{\rm bar} < 0.4$) mostly have $\beta <$0.5 and $\alpha$ $\leq$1.0, whereas the exponential bars have $\beta >$0.5. Therefore both {\ser} and modified Ferrer profiles can describe the degree of flatness and our results from the previous section is insensitive to the choice of fitting functions.


\section{The Global Shape of Bars}   
\label{sec:bar_shape}
\begin{figure}
\begin{center}
\includegraphics[width=8cm]{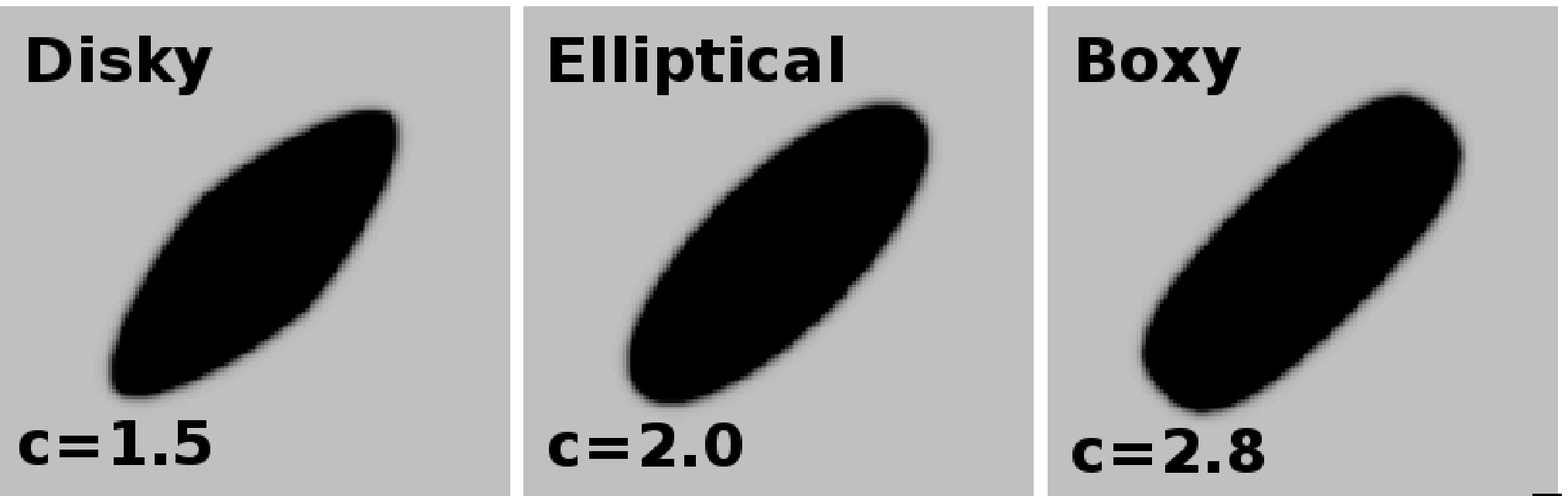}
\caption{Morphology of bars that are simulated with {\sc BUDDA}. From left to right, we plot disky, elliptical, and boxy shape of bars. The shape parameter `c' from the Equation \ref{eq_bar_shape} is written on the bottom left corner for each panel. All bars have ellipticity (1-b/a) of 0.65.}
\label{fig:bar_shape3}
\end{center}
\end{figure}
\begin{figure}
\begin{center}
\includegraphics[width=8cm]{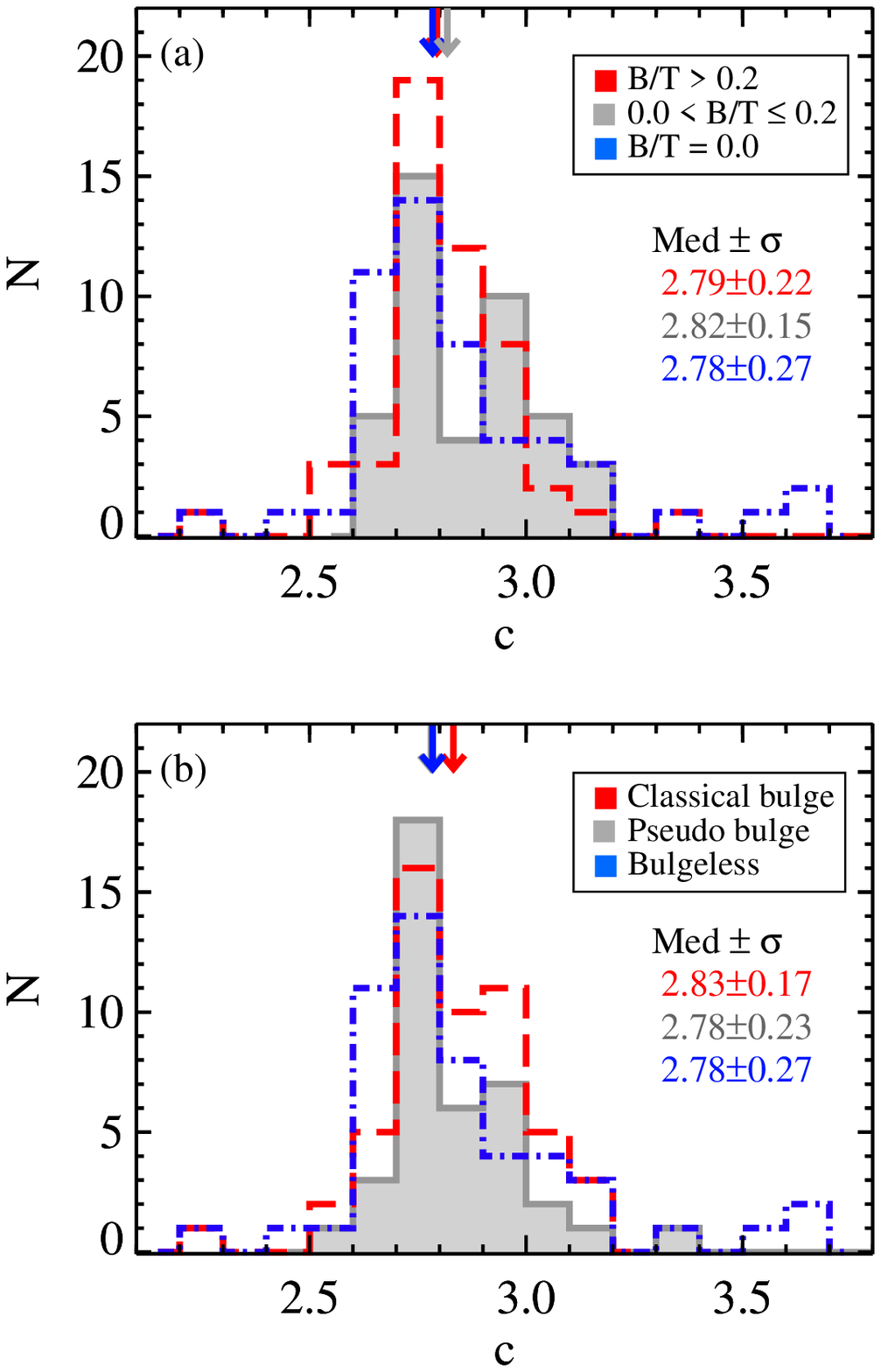}
\caption{Distribution of the bar shape parameter (c from the Equation \ref{eq_bar_shape}) in the generalized ellipse  fit for different types of galaxies. Median shape parameters for each group are plotted with downward arrows. 
(a) Galaxies are divided by B/T:  B/T $>$ 0.2 (dashed lines), 0.0 $<$ B/T $\leq$ 0.2 (solid lines) , and B/T $=$ 0.0 (dot-dashed lines). 
(b) Galaxies are divided by bulge types: classical bulge galaxies (dashed lines), pseudo bulge galaxies (solid lines), and bulgeless galaxies (dot-dashed lines).}
\label{fig:bar_shape_rst}
\end{center}
\end{figure}
Previous observational studies, which were primarily in the optical, found rectangular-shaped bars in strongly-barred early type galaxies (\citealt{athanassoula_90}),  and galaxies with classical and disky pseudo bulges (\citealt{gadotti_11}).  We now revisit this topic using our MIR data and the detailed decompositions with {\sc BUDDA}.  

Simulations predict that bars are rectangular in shape (\citealt{athanassoula_90, athanassoula_02a, scannapieco_12}).  The manifold theoretical model (\citealt{romero_gomez_06, romero_gomez_07, athanassoula_12, athanassoula_10}), which has been proposed to explain the formation and structure of rings and spiral arms in barred galaxy potentials, argues that orbits of confined chaos play a crucial role in explaining rectangular shape of bars \citep{athanassoula_10}.  In particular the manifold branches produce the building blocks of the rectangular outline of the outer parts of a strong bar.

As discussed in \S \ref{sec:data}, the generalized ellipse equation (Eq.\ref{eq_bar_shape}) has an exponent ``c'', the shape parameter, which can distinguish between a diamond-shaped and disky, a rectangular and boxy, or an elliptical shape. 
To help the reader visualize the different shapes, we show three different shapes of the bar models in Fig~\ref{fig:bar_shape3}. In this study, we model the outermost, global shape of the bar.

In Fig~\ref{fig:bar_shape_rst} we show the distribution of the global bar shape parameter ($c$) of all the bars we fit for this study.
All bars are boxy and have a shape parameter c greater than 2. There are {\em no disky bars} in our sample. We find no significant differences in the global shape of bars with different bulge types -- the median bar shape for all B/T and all three bulge types agrees within one standard deviation of the distributions. 
The Kolmogorov-Smirnov test confirms that we can not rule out a null hypothesis that the bar shapes of groups shown in the Fig~\ref{fig:bar_shape_rst} are drawn from the same parent population with the smallest probability, P $\sim$ 0.15.

\citet{athanassoula_90} measured shape parameters as a function of bar radius for 12 galaxies.
We compare our global shape parameters from {\sc BUDDA} fit and those of \citet{athanassoula_90} for 5 galaxies in common and find that our global shape parameters correspond to the shape parameter at 0.9 --1.1 of the bar length that \citet{athanassoula_90} obtained.
The robustness of bar shape parameter obtained with {\sc budda} has been tested in \citet{gadotti_08}, but we also have tested on artificial galaxies of various effective radii and {\ser} indices. We find that the shape parameters estimated from {\sc budda} agree within 10\% of the input value.

Simulations agree with the data that bars are boxy (e.g., \citealt{athanassoula_90, athanassoula_02a, athanassoula_10, scannapieco_12}). 
In addition to what \citet{athanassoula_90} found for strongly barred early type galaxies and \cite{gadotti_11} found for classical and disky pseudo bulge galaxies, we find that bars are rectangular even in late type disk galaxies.  

One caveat is that we fit the bar with a single component. However, bars are known to experience a buckling instability in which the central regions of the bar puff up and extend vertically out from the disk plane. This has been observed as a peanut-shaped or X-shaped feature in some edge-on galaxies (\citealt{jarvis_86}, \citealt{lutticke_00a}). Even in some moderately face-on galaxies, central parts of bars appear boxy over a region where the buckling instability has occurred (\citealt{athanassoula_06, erwin_13}), and this is observed as a bar-lens seen face-on (\citealt{athanassoula_14xx, laurikainen_14}).
Detailed study on the shape of these two different structures of the bar will be performed in the near future.

\section{Discussion}
\subsection{Bar Profiles}
Several ideas have been proposed for the different radial light profiles of bars. \citet{combes_93} suggested that bars in early and late type spiral galaxies have resonances at different locations and that this leads to the different light profiles. They also suggested that the flat profiles along the bar in early type disk galaxies are due to an inner Lindblad resonance that is absent in late type galaxies. \citet{quillen_96_astroph} suggested that bars in early type spirals have a flat surface brightness profile because the ellipticities of the main bar orbits change rapidly as they approach co-rotation near the bar end, while bars in late type galaxies show exponential profiles because the resonances are more spread out. \citet{elmegreen_96a} suggest that flat profiles of bar originated from the excessive stars at the bar ends where the orbits crowd near the inner 4:1 resonance. However, late type bars do not show such resonance crowding (\citealt{elmegreen_96a}).  

Some models find that bars end near the 4:1 resonance (\citealt{contopoulos_89, athanassoula_92a, quillen_94, Skokos_02b}) and corotation radius is an upper limit of the bar extent (\citealt{contopoulos_80, sellwood_93}). However, \citet{elmegreen_96a} claim that bars end at no specific resonance, but end in the region covered by many resonances in the range from inner 4:1 to corotation resonance. If the 4:1 and corotation resonance are located close to each other, stellar orbits crowd together between these two resonances. Therefore it may produce a bar with a nearly constant light distribution, i.e., flat bar. Such a crowding of resonances mostly occurs in bright, massive galaxies, i.e., in early type disk galaxies  where the $\Omega - \kappa/2$ has large maxima and therefore bars can be formed with a large pattern speed (\citealt{combes_93}). However in late type disks, $\Omega - \kappa/2$ has low values as a whole, and the bar pattern speed is low, locating the corotation radius further out in the disk. So this crowding of resonances would develop early type bars with flat surface brightness profile. 

N-body simulations of disk galaxies with different central dark matter halo concentration have shown that there are differences in the mass density profiles of bars (\citealt{athanassoula_02a}).  Bars in galaxies with centrally concentrated halos (MH model of \citealt{athanassoula_02a}) show a flat mass density profile followed by a steeply decreasing density profile towards the end of the bar -- this is similar to the \s4g bars in bulge-dominated galaxies in our sample. 
 In the future perhaps high resolution simulation kinematic data can help us test whether the dark matter halos predicted by the simulations are borne out in these galaxies.

Some caveats should be considered. In this study, we assumed that a bar forms in a disk that shows an exponential profile. However, underlying profile of disks may evolve with redshift. For example, half mass radii change with redshift (e.g., \citealt{dutton_11}). Nevertheless, these changes are slow up to $z=1 \sim1.5$, and a recent study (\citealt{kraljic_12}) claims that bars start to form around this epoch. Thus, we expect the impact of intrinsic change of disk profile would be limited at z$< 1 \sim 1.5$.

Disk galaxies at z$> 1 \sim 1.5$  are found to be compact (\citealt{vandokkum_08, vanderwel_11}). Such galaxies exhibit a similar {\ser} index distribution with that of the massive ($M_{*}>10^{11} M\sun$) local disk galaxies, though the mean {\ser} index of high-z disk galaxies are a bit larger (\citealt{chevance_12}).  
Most disk galaxies in this mass range today are found to host a bar, thus those compact disk galaxies are expected to form a bar by z$=$0.
If a bar forms in such a compact, exotic galaxy at early epoch, it may take longer to change the bar light profile to a flat one. Depending on the initial condition of disk profiles, this may induce a scatter among high-z progenitors. Thus, if profiles of disks vary among galaxies when bars form, our results on bar profiles might also be affected.

\subsection{Bar Profile Bimodality?}
In \S\ref{sub_bar_prof}, we show that surface brightness profiles of bars change with stellar mass. Among less massive galaxies ($M_{\ast}/${\Msun}$\leq 10^{10}$), most of the galaxies show steeply decreasing exponential profile. On the other hand, flat bars are dominant in massive galaxies ($M_{\ast}/${\Msun}$> 10^{10}$). The transition from exponential to flat profile occurs at $M_{\rm 3.6 \mu m} \sim -20$ mag ($M_{\ast}/${\Msun}$ \sim 10^{10.2}$). 

Interestingly, this is the characteristic mass where the bar fraction is at its minimum (\citealt{nair_10b}) and close to the characteristic mass that corresponds to the rotation veolocity ($V_{\rm c} \sim$120 $km~s^{\rm -1}$, see \citealt{comeron_14bxx}) where vertical structures of ISM traced by dust morphology show a transition (\citealt{dalcanton_04, yoachim_06}). 
Also, it is close to the characteristic mass where the mass-to-light ratios of thin and thick disks change (\citealt{comeron_11_thick}), galaxy properties such as age of stellar populations, surface mass density, and concentration show bimodality (\citealt{kauffmann_03b}). Related to this, galaxy color (\citealt{strateva_01, baldry_04}), luminosity (\citealt{balogh_04}), and size (\citealt{shen_03}) are also found to show bimodal distributions.

This suggests that the mechanism that changes bar profiles from exponential to flat may also be associated with global properties of their host galaxies such as galaxy mass, color, size and also distribution of dust lane which is related to disk instability (\citealt{dalcanton_04}). 
\\
\\

\subsection{Are Bars Robust or Recurrent?}
The result that high B/T galaxies do not have an exponential bar implies that such galaxies did not dissolve a pre-existing bar and did not build a new one. 
Some studies have argued that the central mass concentration can destroy bars (e.g., \citealt{hasan_90, hasan_93,norman_96}). 
If bars in such high B/T galaxies are destroyed and formed again, at least some bars should exhibit an exponential bar profile even among high B/T galaxies. However, we do not see any in our study.
Simulations find that with the central mass concentration, the strength of the bar decreases (e.g., \citealt{shen_04, athanassoula_05b}). However, to be able to completely destroy a bar, the mass concentration has to be either highly concentrated whose scale length is less than a few pc or massive enough at least several percent of the disk mass (\citealt{shen_04, athanassoula_05b, debattista_06}). 
For the same central mass concentration, \citet{athanassoula_05b} find that bars with a exponential surface density profile can be dissolved, while strong bars with a flat surface density profile witness only a decrease of their strength. 
This is consistent with our results that higher B/T galaxies only shows flat bars, and this implies that at least bars in galaxies with a big central bulge (high B/T) are {\it not} recurrent.

\subsection{The Invariant Bar Shape and the Bar Profile: An Indicator of Bar Age?}
We find that the global shapes of bars does not vary across galaxy mass or bulge dominance. This suggests that either i) through the bar formation phase and secular evolutionary phase, the global shapes of bars do not change much, or ii) the global shapes of bars evolve to have similar shapes.
However the change in the surface brightness profile from exponential to flat suggests that there is evolution in the number of stars that are trapped in the bar orbits (\citealt{sellwood_93, sellwood_13_rev, athanassoula_03, athanassoula_13_book}).
Because bars are formed out of disk material, we can assume that the light profile of the bar would be exponential-like when they just formed. However, as a galaxy ages and the bar evolves, the galaxy would have enough time to experience resonance crowding (\citealt{combes_93}) and trap stars into the bar orbit and thus the bar becomes longer and stronger (\citealt{sellwood_93, athanassoula_02a, athanassoula_13_book, sellwood_13_rev}). This will lead bars to have a flat profile. Therefore, it is reasonable that the light profile of the bar will change from initial exponential profile to a more flat profile. 

This fits in well with our understanding of the cosmological evolution of disks and bars. The fraction of bars in $L^{*}$ and brighter galaxies are found to evolve such that it increases from z=0.85 to the present day (\citealt{abraham_99, jogee_04, sheth_08, cameron_10, kraljic_12, melvin_14}).
Moreover, these studies have shown that the bar fraction is the highest in the most massive, bulge-dominated, red galaxies at high redshifts with little evolution in this population over the last 7 Gyrs of cosmic time. 
This is perfectly consistent with this study. 
We find that massive, bulge-dominated galaxies have flat bars as one would expect if these bars have been in existence for several Gyrs and dynamically more evolved. Interestingly, this is also consistent with the result of \citet{holwerda_12b} that dust lanes in edge-on galaxies have been in existence since z $\sim$ 0.8 in massive galaxies. This can be interpreted such that massive galaxies had enough time to dynamically evolve so that cold ISM can sink into the galactic plane to form dust lane (\citealt{dalcanton_04}). 

In low-mass, disk-dominated blue galaxies, studies show that the fraction of bars evolves gradually, increasing the present day fraction over time (\citealt{abraham_99,sheth_08, cameron_10, melvin_14}). This means that some low mass systems had their bars early but more and more of them acquired bars over the last 7 Gyrs of evolutionary time. Therefore, today we might expect a larger spread in the bar profile with more exponential profiles in the late type galaxies, as we see from the analysis presented here.  
Thus it stands to reason that if we could ascertain bar profile evolution with time, then we might be able to age-date a bar.  However the rate of capturing stars onto bar orbits may itself evolve in time due to minor mergers, star formation and other processes so the age indicator may still remain elusive.


There have been several studies to infer various age of bars -- using vertical velocity dispersion ($\sigma_{z}$) of bar (\citealt{gadotti_05}), comparing gas mass with gas accretion rate in the bar region (\citealt{elmegreen_09}), age of stellar populations in bars (\citealt{sanchez_blazquez_11}). However, we should be careful what we refer to the bar age. As bars are built out of disk stars and gas, the time elapsed since the formation of bar structure is not necessarily the same as the age of stellar populations that make up the bar.

Although the exact time since the bar formation cannot currently be easily determined, it is certainly a very important parameter to measure in order to understand the impact of bars in galaxy evolution. We expect that the flatness of the bar profile, combined with theoretical work, can be helpful in devising a way to measure the ``dynamical age'' of the bar. We expect that, in general, more dynamically evolved bars should have a flatter surface brightness profile, and have presented the observational evidence that this is indeed the case.

\subsection{Necessity for Cosmic Gas Accretion?}
Previous studies have shown that the formation epoch of bars is correlated with the galaxy host properties -- in other words, the more massive, bulge-dominated, early-type disk galaxies formed their bars early (at z $\gtrsim$ 1, \citealt{sheth_08, cameron_10, kraljic_12, melvin_14}).  
Once formed, bars are resilient and are not likely destroyed easily (e.g. \citealt{athanassoula_03, athanassoula_13, romano_diaz_08b}) without major mergers. 

At the same time, many barred spiral galaxies contain molecular gas within the bar radius (e.g., \citealt{sakamoto_99, sheth_02, sheth_05}). This is surprising given that bars in early type galaxies have likely been in existence since z ${\sim}$ 1 ($\sim$8 Gyrs). 
\citet{elmegreen_09} have argued that the ratio of the gas mass divided by the gas accretion rate may be used as an age for the bar. 
Typical disk gas surface densities in nearby spiral disks is $\sim$ 5--10 {\Msun} pc$^{-2}$ (\citealt{young_95, sheth_05}). Typical bar radius is $\sim$ 2.5 kpc (\citealt{erwin_05_bar}). Therefore one expects the total gas mass inside the bar co-rotation of $\sim 10^9$ {\Msun}. 
Then all the gas within the bar corotation radius should be deposited into the inner Lindblad resonance region within ${\sim}$1 Gyr if we assume the gas inflow rate of 1{\Msun} $yr^{-1}$. 
Indeed, with a detailed estimate, \citet{elmegreen_09} argue that the bar in NGC 1365 is not much older than 1--2 Gyrs.

For barred galaxies, gas within the bar corotation radius is driven inwards and outside the corotation radius is driven outwards. Thus in-plane accretion can only come from around the end of the bar region except for a special case\footnote{If the spiral arm and bar are connected and their pattern speeds match, gas could be transported inwards from the outer disk via spiral arms and drive episodic fueling from outer disk (\citealt{sheth_05}).}. Then how can there be molecular gas in these bars if we have a finite reservoir of gas and a star formation rate in the bar region? 
We argue that the gas might have been replenished via cosmic gas accretion (e.g., \citealt{sancisi_08, dekel_09, fraternali_06, fraternali_14, silk_12, combes_14}) for these barred galaxies.
In general, gas in spiral galaxies can be replenished from outer disks where there is lots of gas that can come inward from spiral torques. However, for barred spiral galaxies the amount of gas that can be transported inward is limited. Therefore, we indeed need cosmic gas accretion to sustain bars and allow them to grow slowly over time.
This is consistent with the results from cosmological simulations of \citet{kraljic_12} that expects slow emergence of bars from z $\sim$ 1, and the slow down of the bar growth in the presence of gas (\citealt{athanassoula_13}).
Lastly, such cosmic accretion might be the origin of the gas that bring galaxies to evolve and renew bars (\citealt{bournaud_02, block_02}).

\section{Summary and Conclusion}
We make use of 2D image decompositions on 144 barred galaxies of Hubble types from SB0 to SBdm using 3.6 \mus images drawn from the \s4g. We investigate the structural properties of bars, in particular radial light profiles of bars and 2D global shapes of bars. 
We summarize our results as follows.

\begin{itemize}
\item 
We quantify the surface brightness profile of bars by fitting bar isophotes with the {\ser} profile. We find that massive, higher B/T, and classical bulge galaxies tend to have flat bars, while less massive and bulgeless galaxies tend to show steeply decreasing exponential bar profiles. We find that whenever there is a bulge, galaxies tends to have flat bars.

\item 
We model the global shape of bars with generalized ellipses. All bars are found to be rectangular-like, i.e., boxy. There are no significant differences in the shape of bars among galaxies. This implies that as bars evolve, light profiles of bars change from exponential to flat, although their outermost shapes remain boxy.

\item
We conjecture that at earlier evolutionary stages, the bar profile resembles that of the disk that shows an exponential profile ($<n_{bar}> \sim$1). But as galaxies evolve, bars become stronger and this leads to the developement of flatter profiles. 
In this regard, our findings are consistent with the cosmological evolution of barred galaxies which predict that more massive, bulge-dominated, red disk galaxies formed their bars first, and thus had enough time for their bars to evolve towards flat profiles. Combined with theoretical works and simulations, we will be able to diagnose dynamical status of bars using light profile of bars.

\item
Cosmic gas accretion is strongly required to explain the presence of gas and star formation within bar region for barred galaxies that have been in existence for more than their gas consumption time scale ($\sim$ 1--2 Gyrs).
\end{itemize}

\acknowledgments
{\it Facilities:} \facility{ The {\em{Spitzer}} Space Telescope}
\\
\\
We thank Frederic Bournaud, the referee, for insightful comments that greatly improved this paper.
The authors thank the \s4g team for their effort in this project.
T.K. is grateful to Woong-Tae Kim and his research group for their helpful comments and discussions.
T.K., K.S., J.-C.M.-M., and T.M. acknowledge support from the National Radio Astronomy Observatory.  
The National Radio Astronomy Observatory is a facility of the National Science Foundation operated under cooperative agreement by Associated Universities, Inc.
We are grateful for the support from NASA JPL/Spitzer grant RSA 1374189 provided for the {\s4g} project.
T.K. and M.G.L. were supported by the National Research Foundation of Korea (NRF) grant funded by the Korea Government (MEST) (No. 2012R1A4A1028713).
T.K. acknowledges the support from ESO for the studentship in 2011--2012.
D.A.G. thanks funding under the Marie Curie Actions of the European Commission (FP7-COFUND).
E.A. and A.B. thank the Centre National d'Etudes Spatiales for financial support. We acknowledge financial support from the People Programme (Marie Curie Actions) of the European Union's FP7/2007-2013/ to the DAGAL network under REA grant agreement No. PITN-GA- 2011-289313.
L.C.H acknowledges support by the Chinese Academy of Science through grant No. XDB09030102 (Emergence of Cosmological Structures) from the Strategic Priority Research Program and by the National Natural Science Foundation of China through grant No. 11473002.
This research is based on observations and archival data made with the {\em{Spitzer}} Space Telescope, and made use of  the NASA/IPAC Extragalactic Database (NED) which are operated by the Jet Propulsion Laboratory, California Institute of Technology under a contract with National Aeronautics and Space Administration (NASA). We acknowledge the usage of the HyperLeda database (http://leda.univ-lyon1.fr).
\\
\\
\\
{$^1$}{Astronomy Program, Department of Physics and Astronomy, Seoul National University, Seoul 151-742, Korea} {Email: thkim@astro.snu.ac.kr}\\
{$^2$}{National Radio Astronomy Observatory/NAASC, 520 Edgemont Road, Charlottesville, VA 22903, USA}\\
{$^3$}{European Southern Observatory, Casilla 19001, Santiago 19, Chile}\\
{$^4$}{University of Arizona, 933 N. Cherry Ave, Tucson, AZ 85721, USA}\\
{$^5$}{IBM Research Division, T.J. Watson Research Center, Yorktown Hts., NY 10598, USA}\\
{$^6$}{Aix Marseille Universit{\'e}, CNRS, LAM (Laboratoire d'Astrophysique de Marseille) UMR 7326, 13388 Marseille, France}\\
{$^7$}{European Space Agency, ESTEC, Keplerlaan 1, 2200-AG, Noordwijk, The Netherlands}\\
{$^8$}{Kavli Institute for Astronomy and Astrophysics, Peking University, Beijing 100871, China}\\
{$^9$}{Department of Astronomy, School of Physics, Peking University, Beijing 100871, China}\\
{$^{10}$}{Division of Astronomy, Department of Physical Sciences, University of Oulu, Oulu, FIN-90014, Finland}\\
{$^{11}$}{Finnish Centre of Astronomy with ESO (FINCA), University of Turku, V{\"a}is{\"a}l{\"a}ntie 20, FI-21500, Piikki{\"o}, Finland}\\
{$^{12}$}{Instituto de Astrof\'\i sica de Canarias, E-38200 La Laguna, Tenerife, Spain}\\
{$^{13}$}{Departamento de Astrof\'\i sica, Universidad de La Laguna, E-38205 La Laguna, Tenerife, Spain}\\
{$^{14}$}{MMTO, University of Arizona, 933 North Cherry Avenue, Tucson, AZ 85721, USA}\\
{$^{15}$}{Department of Physics and Astronomy, University of Alabama, Box 870324, Tuscaloosa, AL 35487, USA}\\
{$^{16}$}{Korea Astronomy and Space Science Institute, Daejeon 305-348, Republic of Korea}\\
{$^{17}$}{The Observatories of the Carnegie Institution of Washington, 813 Santa Barbara Street, Pasadena, CA 91101, USA}\\
{$^{18}$}{Universidade Federal do Rio de Janeiro, Observat{\'o}rio do Valongo, Ladeira Pedro Ant{\^{o}}nio, 43, CEP 20080-090, Rio de Janeiro, Brazil}\\
{$^{19}$}{Space Telescope Science Institute, 3700 San Martin Drive, Baltimore, MD 21218, USA}\\
{$^{20}$}{South African Astronomical Observatory, Observatory, 7935 Cape Town, South Africa}\\
{$^{21}$}{Departamento de Astrof\'\i sica, Universidad Complutense de Madrid, Madrid 28040, Spain}\\
{$^{22}$}{Florida Institute of Technology, Melbourne, FL 32901, USA}
\\
\\
\bibliography{tkim_bar}

\end{document}